\documentclass[acmtog, authorversion]{acmart}

\AtBeginDocument{%
  \providecommand\BibTeX{{%
    \normalfont B\kern-0.5em{\scshape i\kern-0.25em b}\kern-0.8em\TeX}}}

\setcopyright{rightsretained}
\copyrightyear{2021}
\acmYear{2021}
\acmDOI{}
\acmISBN{} 
\acmConference[Virtual '21]{Virtual '21: ACM SIGKDD International Conference on Knowledge Discovery & Data Mining Workshop}{August 14-18, 2021}{Virtual '21}
\acmBooktitle{the KDD Workshop on Data-driven Humanitarian Mapping held with the 27th ACM SIGKDD Conference on Knowledge Discovery and Data Mining (KDD ’21), August 14, 2021.}
\begin{document}

\title{Sensing and Mapping for Better Roads: Initial Plan for Using Federated Learning and Implementing a Digital Twin to Identify the Road Conditions in a Developing Country - Sri Lanka}

% \affiliation{%

% \affiliation{%

% \affiliation{%

\author{Thilanka Munasinghe}
\affiliation{%
 \institution{School of Science, Rensselaer Polytechnic Institute}
 \streetaddress{110 8th St}
 \city{Troy}
 \state{New York}
 \country{USA}}

\author{HR Pasindu}
\affiliation{%
  \institution{Department of Civil Engineering, University of Moratuwa}
  \streetaddress{Katubedda}
  \city{Moratuwa}
  \state{Western Province}
  \country{Sri Lanka}}

% \affiliation{%

% \affiliation{%

% \affiliation{%

\renewcommand{\shortauthors}{Munasinghe, et al.}

\begin{abstract}
We propose how a developing country like Sri Lanka can benefit from privacy-enabled machine learning techniques such as Federated Learning to detect road conditions using crowd-sourced data collection and proposed the idea of implementing a Digital Twin for the national road system in Sri Lanka. Developing countries such as Sri Lanka are far behind in implementing smart road systems and smart cities compared to the developed countries. The proposed work discussed in this paper matches the UN Sustainable Development Goal (SDG) 9: "Build Resilient Infrastructure, Promote Inclusive and Sustainable Industrialization and Foster Innovation".  Our proposed work discusses how the government and private sector vehicles that conduct routine trips to collect crowd-sourced data using smartphone devices to identify the road conditions and detect where the potholes, surface unevenness (roughness), and other major distresses are located on the roads. We explore Mobile Edge Computing (MEC) techniques that can bring machine learning intelligence closer to the edge devices where produced data is stored and show how the applications of Federated Learning can be made to detect and improve road conditions. During the second phase of this study, we plan to implement a Digital Twin for the road system in Sri Lanka. We intend to use data provided by both Dedicated and Non-Dedicated systems in the proposed Digital Twin for the road system.  As of writing this paper, and best to our knowledge, there is no Digital Twin system implemented for roads and other infrastructure systems in Sri Lanka. The proposed Digital Twin will be one of the first implementations of such systems in Sri Lanka. Lessons learned from this pilot project will benefit other developing countries who wish to follow the same path and make a data-driven decisions. Additionally, our intended work can be used as a blueprint for those countries planning to implement such systems.
\end{abstract}




%   <concept>
%       <concept_id>10010520.10010553.10003238</concept_id>
%       <concept_desc>Computer systems organization~Sensor networks</concept_desc>
%       <concept_significance>500</concept_significance>
%       </concept>
%  </ccs2012>

% \ccsdesc[500]{Computer systems organization~Sensor networks}






\maketitle

\section{Introduction}

Developing countries have a significantly less budget allocated for road maintenance when compared to developed countries. This increases the life cycle cost of the road assets and reduces the design life. One of the key issues that hinders executing the required maintenance, in addition to budget constraints is the sub-optimal fund allocation which is primarily based on either subjective, ad-hoc decision making. Lack of accurate, timely condition data of the road assets is one of the constraints faced by road agencies when adopting quantitative techniques to optimize maintenance strategies for roads.  This proposed work in line with the UN Sustainable Development Goals (SDG), Goal 9: Building Resilient Infrastructure \cite{UNSDG, UNSDG2}. Optimizing road asset maintenance ensures asset design life is maximized and the infrastructure is resilient to any changes in the environment. Leveraging on technology allows the agencies to make sure road condition data collection can be carried out efficiently as well as in a cost-effective manner.
In this vision paper, we propose how a developing country like Sri Lanka can benefit from privacy-enabled machine learning techniques such as Federated Learning to detect road conditions using crowd-sourced data collection and proposed idea of implementing a Digital Twin for the national road system in Sri Lanka.

\subsection{Road conditions in Sri Lanka/ Sri Lanka as a case study}

According to the Road Development Authority (RDA), Sri Lanka has one of the highest road densities in the region, with 1.74 km/sq.km. The national road network length is approximately 12,500 km in length; however, the majority of the road network is part of the provincial and local roads which are around 18,900 km and 88,200 km respectively \cite{NationalRoadMasterPlanSL}. The key challenges faced by the road agencies is as follows,
1. Ensure the national roads are maintained at the required pavement (sidewalk) condition.
2. Forecast budget requirement for different maintenance strategies
3. Allocate funds for provincial and local roads based on their condition.
Budget Allocation:
The annual maintenance need for the national roads is around USD 80 million. An efficient maintenance management system has the potential to save around 20-30 percent of the maintenance cost. When we factor in the maintenance requirements of the other roads in the network, this is a massive burden to the economy of the country, which has a GDP of USD 80 billion \cite{NationalRoadMasterPlanSL}. Thus, there is a need to improve the maintenance strategies to ensure optimal resource allocation for maintenance and also to forecast future maintenance requirements for financial planning. Pavement (sidewalk) condition monitoring and evaluation have a significant role to play in this scenario. It can be the basis for the road asset management system, which will provide the most cost-effective maintenance strategies. The proposed method aims to assist in this endeavor by making the pavement data collection process more efficient.

\section{Related Work}

Cities, roads and other infrastructure are becoming more and more digitally connected due to the abundance of connected devices these days. Gerhard et.al explains how an automated data collecting process will play a leading role in smart cities and also discuss some of the challenges which are technological and societal in nature in ubiquitous sensing in urban environments \cite{hancke2013role}. Well managed and maintained network of roads are essential for smart cities.  Okai et.al elaborates on the benefits and challenges of sensing in smart cities and their foreseen future and what features will characterize a smart city \cite{okai2018smart}. Advancements in wireless communication and computing technologies have propelled the use of cloud computing, the Internet of Things (IoT), and intelligent technologies to the next level \cite{ali2019security, jiang2020federated}. The essence of IoT is the connection of devices through the internet, which enables connectivity among things (devices) or services and people \cite{vermesan2011internet, ngu2016iot, al2020survey}. Capponi et al. discuss their survey on opportunities, challenges, and solutions on Mobile crowdsensing (MCS) systems and explain how a MCS system can rely on a contribution by the participation of many mobile devices or a crowd \cite{capponi2019survey}. 

The concept of Sensing as a Service (S2aaS) allows data from sensors connected to embedded devices, and it was influenced by the cloud computing term "Every Thing as a Service". Alarbi et.al explain their paper "Sensing as a Service Middleware Architecture",on how middleware enables to access the data that generated by IoT devices which are own by other entities \cite{alarbi2018sensing}.
Various applications such as monitoring road conditions can benefit from Sensing as a Service. \cite{lau2017sensor} Smart cities and other infrastructure projects can benefit from Sensing as a Service when implementing self-monitoring systems \cite{li2019survey}.
Zaslavsky et al., in their paper on "Sensing as a Service and Big Data", describe how the federating sensor networks can be applied and their challenges that faced during cloud-based management, storing, and processing sensor data \cite{zaslavsky2013sensing}.

Brisimi et al., in their paper " Sensing and Classifying Roadway Obstacles in Smart Cities: The Street Bump System," have introduced a method for detecting road obstacles by using smartphone sensing information. \cite{brisimi2016sensing}.
Li et al. in their survey on Federated Learning Systems: Vision, Hype and Reality for Data Privacy and Protection have conducted a comprehensive review on federated learning systems, and they provide a thorough categorization based on different aspects of federated learning, which are: data distribution, motivation for privacy mechanism, machine learning model and communication architecture \cite{li2019survey}.

\section{Data Collection}

Smart phone accelerometer and GPS data can be used to detect where the potholes are located \cite{mednis2011real, sattar2018road}.
Our proposed work tends to use the government and private sector vehicles that conduct routine trips to identify the road conditions and detect where the potholes, surface unevenness (roughness), and other major distresses that are located on the roads.  These vehicles can be government postal service vehicles and private-sector delivery service vehicles willing to install the application on the mobile phone device mounted in the vehicles. Instead of dedicating a special vehicle to collect the data, we wish to use the crowd-sourced mechanism to collect the data. By crowd-sourcing to collect the needed data, the government can save a significant amount of money, and the saved resources can be applied to other research and development work which is vital for developing countries \cite{shu2017mobile}. An additional benefit of crowd-sourcing is that it is possible to obtained rich data sets that frequently get updated. As we approach this crowd-sourced data collection, it is essential to understand the data provider's privacy. This is the main reason that we intend to use privacy-preserving federated learning \cite{GoogleFederateLearningBlog}.

\section{Proposed Methodology}

There are two types of sensing systems; (1) Dedicated and (2) Non-Dedicated \cite{jiang2020federated}. Dedicated sensing systems are a network of sensors set up permanently and expressly for a particular application. An example of Non-Dedicated sensing is when mobile phone devices being used for sensing purposes. They are considered as Non-Dedicated sensing because end-user can dictate when to use the application and collect the data \cite{jiang2020federated, habibzadeh2018soft, habibzadeh2017large}. 
During the initial phase of our proposed study, we intend to use a Non-Dedicated sensing approach using mobile phone devices. As our work progresses to the next phase of the project, we intend to implement both Dedicated and Non-Dedicated sensing systems to obtain rich datasets. Since there is an abundance of smart mobile phone devices and the ease of utilizing an App, using these non-dedicated sensing via mobile phones will add cost-benefit to the research project by saving money.  It is possible to attract users by providing an incentive to the users to participate by pay-per-use model. Providing a financial incentive and using a pay-per-use model can alleviate a massive initial investment cost for the project's stakeholders \cite{misra2014theoretical}. We foresee some challenges when using mobile devices for incentive-based crowd-sourced data; there are potential malicious attacks, threats, and exposure to privacy-leaks of the users \cite{lin2017frameworks, wu2017dynamic, jin2016inception}.

\subsection{Application of Federated Learning}

As we proposed a crowd-sourced, Non-Dedicated sensing mechanism of collecting data utilizing smart mobile phones, we propose using a privacy-preserving federated learning system to collect and analyze the data \cite{GoogleFederateLearningBlog, li2019survey}. 
The concept of Federated Learning was first introduced by Google in year 2016. \cite{konevcny2016federated}
Goal of the federated learning is to use edge devices to conduct stated of the art machine learning without centralizing data and preserving the privacy \cite{GoogleFederateLearningBlog, li2020federated}. Lim et.al explains how Mobile Edge Computing (MEC) techniques that being proposed to bring machine learning intelligence closer to the edge devices where the data is produced and stored and show how the applications of Federated Learning for mobile edge network optimization can be made \cite{lim2020federated}.  As stated in the related work section, based on what other researcher have done, it is evident to us that use of Mobile Edge Computing (MEC) supported by Federated Learning approach the most viable solution to identify road conditions.

In general, federated learning is classified into three types: sample-based (horizontal) federate learning, feature-based (verticle) federated learning, and federated transfer learning \cite{yang2019federated, lim2020federated}. As of writing this vision paper, authors are researching the current state-of-the-art federated learning type that can be applied in this proposed study. As an initial step, we tend to apply a sample-based (horizontal) federated learning technique where we intend to train our global model in the server and deploy the trained global model to the edge devices. Then we plan to optimize the model from the edge devices.

In order to apply federated learning successfully, it is necessary to send updates to the server. There are two well-known methods:  Federated Stochastic Gradient Descent (FedSGD) and Federated Averaging (FedAvg). FedSGD is a technique that was inspired by well known statistical optimization method of Stochastic Gradient Descent  
\cite{yang2019federated, li2018federated, mcmahan2017communication}. 
In contrast to FedSGD, Federated Averaging (FedAvg) is a heuristic method based on averaging local Stochastic Gradient Descent (SGD) \cite{mcmahan2017communication} which adds more computation to each client.  In addition to the two techniques stated above, authors are exploring a technique called FedProx, described by Li et.al; in their paper on "Federated optimization in heterogeneous networks." FedProx is a state-of-the-art method for federated learning which is a generalization, and re-parametrization of the technique of FedAvg method \cite{li2018federated}, and authors are currently exploring those techniques and how it can be applied in this proposed study.

\subsection{Challenges with Federated Learning and Data Collections and Implementation}
When using sensors on mobile devices, such as accelerometers, to detect potholes or surface unevenness may result in false positives and false negatives. 
In addition to possible false positive and false negative results, authors foresee several challenges when preserving privacy due to possible attacks. Li et.al describes possible challenges in their paper on "Preserving Data Privacy via Federated
Learning: Challenges and Solutions" related to possible attacks \cite{li2020preserving} that can compromise the privacy. Those attacks can be inference attacks or model poisoning attacks. An inference attack is an attack that can infer some sensitive information to which has not been granted access \cite{pyrgelis2018location, li2020preserving, orekondy2018gradient, melis2019exploiting}. A poisoning attack is an attack when adversaries intentionally inject bad data (incorrect data) into the model’s training pool, allowing and making the model to learn that it should not be learned \cite{bagdasaryan2020backdoor, li2020preserving}. Currently, as of writing this vision paper, authors are investigating the latest and state-of-the-art techniques that has been used by other researcher in the community to apply in this study to mitigate and address the challenges.   

Secondly, as we progress, during the second phase of the proposed study, we intend to utilize an additional pavement performance indicators such as vehicle mounted laser sensors, or permanent sensors on the pavement will support implementing a Digital Twin for the road system in Sri Lanka. We intend to use data provided by both Dedicated and Non-Dedicated systems in the proposed Digital Twin for the road system.  As of writing this paper, and best to our knowledge, there is no Digital Twin system implemented for roads and other infrastructure systems in Sri Lanka. The proposed Digital Twin will be one of the first implementations of such systems in Sri Lanka.

\subsection{Implementation of a Digital Twin}

Real world physical object or a system can be represent digitally using a "Digital Twin". Digital Twin allows the data scientist, researchers and IT professionals to optimize the complex connected systems to optimize and tune the system to it's peak efficiency \cite{WhatIsDigitalTwin}. 

Digital twin consists of three key parts 
(1) Data collection module
(2) Performance prediction module
(3) Parametric analysis module
% \cite{https://doi.org/10.1155/2020/8824135}
\cite{yu2020prediction}
The data collection module includes pavement performance, maintenance records,highway structure, traffic flow, and environment. Pavement performance data come from pavement performance indicators, such as vehicle-mounted laser sensors and high precision accelerators. These are often costly equipment and not readily available for the entire road network. Therefore the performance data availability is low. By adopting the proposed approach in the study, this constraint can be alleviated. Moreover, pavement performance prediction modelling can also be improved from the proposed study as it can leverage on machine learning techniques to evaluate how the pavement performance indicators have varied in different pavement segments based on the traffic flow, environment, pavement age etc. which are the basis for performance prediction.

When it comes to regularly monitor the conditions of an infrastructure such as roads, it is extremely useful for the engineers and road construction personals to have a digital twin to monitor and conduct critical maintenance decision on elements either needed to be repaired or replaced in order to make the road and other infrastructures are safe and also to obtain the maximum durability. Additionally, a digital twin allows us to make intelligent maintenance decisions and monitor infrastructure performances under various service conditions and generate simulations by mimicking the real-world scenarios \cite{WhatIsDigitalTwin2}. In our proposed work, we are planning to test how the proposed methodology can be adapted in the development of a digital twin for the national road system using the connected mobile devices that provide the data through crowd-sourcing. As digital twin connects the physical and virtual world, the collected data can be stored in locally decentralized or centralized in a cloud platform. During this study we explore the possibilities on how to implement a digital twin for the road system in Sri Lanka.

%\cite{LearningObjectiveDataAnalytics2}

% We intend to generate a simulated environment based on the input data.  //may be better to omit this sentence, we can only facilitate part 1,2 of the digital twin, the third requires extensive data inputs, which may not be a possibility//

\section{Overall Vision}
To summarize this short paper, our vision is to share the initial plan of using privacy-preserving federated learning to detect road conditions and implementing a Digital Twin for a road system in Sri Lanka. Developing countries such as Sri Lanka are far behind in implementing smart road systems and smart cities compared to the developed countries. This proposed work matches the UN Sustainable Development Goal (SDG) 9: "Build Resilient Infrastructure, Promote Inclusive and Sustainable Industrialization and Foster Innovation" \cite{UNSDG}. Lessons learned from this pilot project will benefit other developing countries who wish to follow the same path and make data-driven decision, additionally our intended work can be used as a blueprint for those countries planning to implement such systems.

\bibliographystyle{ACM-Reference-Format}
\bibliography{sample-base}

\appendix









\end{document}